\documentclass[aps,preprint,preprintnumbers,amsmath,amssymb]{revtex4}
\usepackage{amssymb}
\usepackage{graphicx}
\usepackage{dcolumn}
\usepackage{bm}
\usepackage{color}

\makeatletter
\newcommand{\rmnum}[1]{\romannumeral #1}
\newcommand{\Rmnum}[1]{\expandafter\@slowromancap\romannumeral #1@}

\begin{document}

\title{Pressure-tuned quantum criticality in the antiferromagnetic Kondo semimetal CeNi$_{2-\delta}$As$_2$}

\author{Yongkang Luo$^{1}$\footnote[1]{Electronic address: ykluo@lanl.gov}, F. Ronning$^{1}$, N. Wakeham$^{1}$, Xin Lu$^{2}$, Tuson Park$^{3}$, Zhu-an Xu$^{4}$, and J. D. Thompson$^{1}$}

\address{$^1$Los Alamos National Laboratory, Los Alamos, New Mexico 87545, USA;}
\address{$^2$Center for Correlated Matter, Zhejiang University, Hangzhou 310058, China;}
\address{$^3$Department of Physics, Sungkyunkwan University, Suwon 440-746, South Korea;}
\address{$^4$Department of Physics, Zhejiang University, Hangzhou 310027, China.}

\date{\today}

\maketitle

\textbf{The easily tuned balance among competing interactions in Kondo-lattice metals allows access to a zero-temperature, continuous transition between magnetically ordered and disordered phases, a quantum-critical point (QCP). Indeed, these highly correlated electron materials are prototypes for discovering and exploring quantum-critical states. Theoretical models proposed to account for the strange thermodynamic and electrical transport properties that emerge around the QCP of a Kondo lattice assume the presence of an indefinitely large number of itinerant charge carriers. Here, we report a systematic transport and thermodynamic investigation of the Kondo-lattice system CeNi$_{2-\delta}$As$_2$ ($\delta$$\thickapprox$0.28) as its antiferromagnetic order is tuned by pressure and magnetic field to zero-temperature boundaries. These experiments show that the very small but finite carrier density of $\sim$0.032 $e^-$/f.u. in CeNi$_{2-\delta}$As$_2$ leads to unexpected transport signatures of quantum criticality and the delayed development of a fully coherent Kondo lattice state with decreasing temperature. The small carrier density and associated semi-metallicity of this Kondo-lattice material favor an unconventional, local-moment type of quantum criticality and raise the specter of Nozi\`{e}res exhaustion idea that an insufficient number of conduction-electron spins to separately screen local moments requires collective Kondo screening.}\\


\emph{\textbf{SIGNIFICANCE}} An unconventional quantum critical point involves a critical destruction of the Kondo entanglement and an reconstruction of Fermi surface topology. A description of such quantum criticality requires broader experimental basis and theoretical model that includes critical fermionic degrees of freedom. We provided a rare example of peculiar quantum critical behavior in the low carrier density limit. Most significantly, the similarity between our CeNi$_{2-\delta}$As$_2$ and the well known quantum-critical Kondo-lattice system CeCu$_{6-x}$Au$_x$ indicates that a condition favorable for the unconventional quantum criticality is a ``small" Fermi volume which disfavors the conventional Hertz-Millis type spin-density-wave criticality. This insight provides new guidance for where new examples of unconventional quantum criticality could be found.\\

During the past decade or so, particular interest in Kondo-lattice systems has focused on those in which a moderate hybridization ($J_{fc}$) between magnetic $f$-electrons and a sea of itinerant charge carriers allows their tuning by a non-thermal control parameter to a quantum-critical point (QCP) where non-Fermi-liquid (NFL) signatures appear in transport and thermodynamic properties\cite{Doniach}. Though several models of quantum criticality have been proposed to account for various NFL properties\cite{HvL,Si-QCP}, a common assumption of these models is that the material is metallic. In these metals, the magnetic order that is tuned towards zero temperature is either of a local-moment type derived from Ruderman-Kittel-Kasuya-Yosida (RKKY) interactions when $J_{fc}$ is relatively weak or a spin-density wave (SDW) instability of a large Fermi surface to which the delocalized $4f$ state contributes when $J_{fc}$ is stronger. An interesting question is what might be expected in a system with a very low carrier density and, additionally, is how the low carrier density might influence the signatures of quantum criticality. A related issue is the nature of the magnetism that is being tuned in such a system. A low carrier density implies a dearth of conduction electrons and, consequently, a small Fermi wave vector. Under these circumstances SDW order is unlikely (but principally not impossible); however, because the RKKY interaction depends on electrons near as well as deeper inside the Fermi sea\cite{Neto-RKKY}, RKKY-mediated order is more favorable.
Additionally, the crossover from a low-temperature Fermi-liquid (FL) state to high-temperature local moment state in a Kondo lattice can be slowed in the low carrier density limit, i.e., so-called protracted Kondo screening\cite{Tahvildar-Protracted,Sarrao-YbXCu4,Lawrence-Crossover}. New materials with tunable long-range magnetism and low carrier density are, therefore, of some interest.

At room temperature, CeNi$_{2-\delta}$As$_2$ ($\delta$$\approx$0.28) crystallizes in the well known ThCr$_2$Si$_2$-type structure (I4/$mmm$, No. 139) but may undergo a very weak orthorhombic distortion at low temperature\cite{Luo-CeNi2As2}. Well below this possible orthorhombic distortion, Ce moments order antiferromagnetically at $T_N$$\approx$5 K\cite{Luo-CeNi2As2,Suzuki-CeNi2X2} with the \textbf{c}-axis being the magnetic easy axis. In the presence of an external magnetic field $\textbf{B}$$\parallel$$\textbf{c}$, the ordered Ce moments undergo a weakly first-order spin-flop transition from an antiferromagnetic (AFM) ground state to a polarized paramagnetic state. A magnetic structure was proposed in Ref.~\cite{Luo-CeNi2As2} but remains to be confirmed; nevertheless, the existence of a spin-flop transition below $T_N$ suggests that the order is of a local-moment type, which is also consistent with a modest zero-temperature Sommerfeld coefficient of 65 mJ/mol$\cdot$K$^2$ estimated by extrapolating the specific heat divided by temperature from above $T_N$\cite{Luo-CeNi2As2}. The magnetic entropy below $T_N$ of approximately 0.6$R\ln2$ indicates magnetic order in a crystalline electric field (CEF) doublet ground state and some $f$-$c$ hybridization. Herein, we report the effects of hydrostatic pressure and applied magnetic field on the transport and thermodynamic properties of CeNi$_{2-\delta}$As$_2$. At atmospheric pressure, a pronounced anomalous Hall effect (AHE) scales well to a magnetization anomaly at the spin-flop, and it provides a useful means to track the pressure dependence of magnetic order at low temperatures. Whereas, the normal Hall coefficient confirms a low carrier density in CeNi$_{2-\delta}$As$_2$. The AFM order is suppressed gradually under pressure and vanishes at $p_c$=2.7 GPa, above which a FL-like $T^2$-resistivity develops. We discuss the possibility of a pressure-driven QCP in this low carrier density Kondo lattice and its relation to field-induced $T$=0 boundary at $B_c$=2.8 T under atmospheric pressure.

\section{Results}

\begin{figure*}[ht]
\begin{center}
\includegraphics[width=18.0cm]{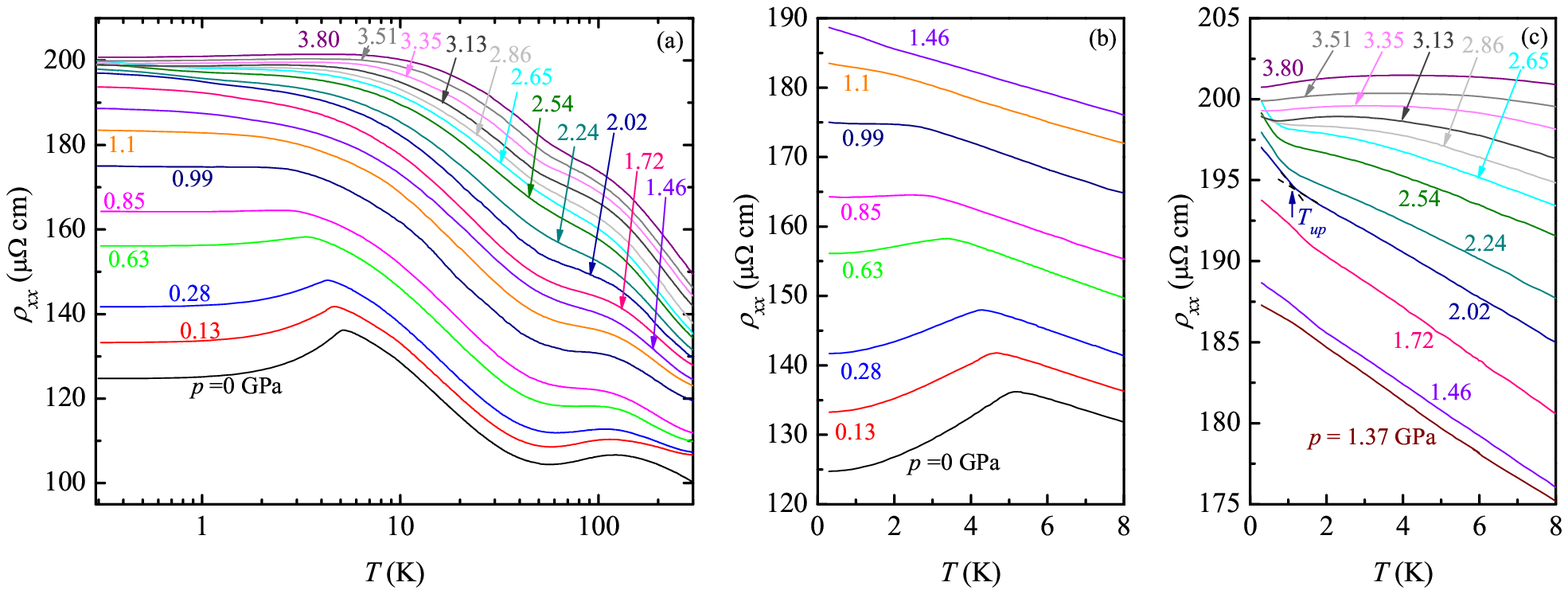}
\caption{Temperature dependence of the in-plane resistivity of CeNi$_{2-\delta}$As$_2$ under various pressures. Panel (a) shows data over the full temperature range; whereas, (b) and (c) display enlarged plots of $\rho_{xx}(T)$ below 8 K. }\label{Fig.1}
\end{center}
\end{figure*}

\begin{figure}
\includegraphics[width=10cm]{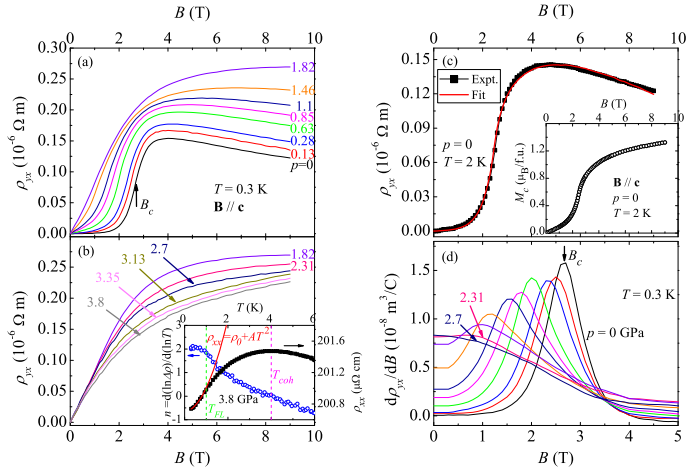}
\caption{Pressure-dependent Hall effect measurments on CeNi$_{2-\delta}$As$_2$. (a) and (b) show the Hall resistivity $\rho_{yx}$ as a function of magnetic field at $T$=0.3 K. (c) shows a fit of $\rho_{yx}(B)$ to Eq.~[\ref{Eq.1}], while the inset is a plot of isothermal magnetization at $T$=2 K with $\textbf{B}$$\parallel$$ \textbf{c}$. (d) shows the derivative of $\rho_{yx} (B)$, and the peak in $d\rho_{yx}/dB$ defines the critical field of a spin-flop transition. Inset to (b) shows a local exponent $n$=$d(\ln\Delta\rho)/d(\ln T)$ (left) and $\rho_{xx}$ (right) as functions of $T$ for $p$=3.8 GPa; $T_{FL}$ and $T_{coh}$ are defined as the temperatures where $n$=1.8 and 0, respectively.}\label{Fig.2}
\end{figure}

The temperature-dependent resistivity $\rho_{xx}(T)$ of CeNi$_{2-\delta}$As$_2$ under various pressures is plotted in Fig.~\ref{Fig.1}(a). At ambient pressure, the large $\rho_{xx}(T)$ increases slowly with decreasing $T$, typical of semi-metallic behavior, and there is a broad hump centered around 110 K. Such a broad hump in resistivity is ascribed to  Kondo scattering on excited CEF levels. Below 50 K, $\rho_{xx}(T)$ increases approximately logarithmically with decreasing $T$, characteristic of Kondo scattering in the CEF doublet ground state. $\rho_{xx}(T)$ develops a sharp peak at $T_N$=5.1 K, indicative of the reduction of spin scattering due to the formation of long-range order of Ce moments. As a function of pressure, five prominent tendencies are apparent. (\rmnum{1}) Overall, the magnitude of resistivity increases with pressure, and except for a narrow pressure range around 3 GPa and at very low temperatures, isobaric curves do not cross. This is not typical of the pressure response of Ce-based Kondo lattice metals. (\rmnum{2}) The hump due to CEF splitting tends to be smeared but its position changes only slightly. (\rmnum{3}) The sharp peak at $T_N$ is suppressed by pressure and is hardly observable when $p$ exceeds 1.1 GPa [Fig.~\ref{Fig.1}(b)]. (\rmnum{4}) With further increasing pressure, an ``inflection'' appears below 2 K and $\rho_{xx}(T)$ turns up [Fig.~\ref{Fig.1}(c)], signalling a further decrease in carrier concentration and/or an increase in scattering rate. It should be pointed out that the evolution from peak to upturn seems continuous. Hall effect and ac heat capacity measurements, discussed below, indicate that in this pressure region Ce moments still order antiferromagnetically at low temperature. (\rmnum{5}) For even higher pressure, the upturn in $\rho_{xx}(T)$ is absent and FL-like behavior with $\rho_{xx}(T)$=$\rho_0$+$\Delta\rho$=$\rho_0$+$AT^2$ is observed below a resistivity maximum at $T_{coh}$ and typical of coherence in a Kondo lattice [data shown in the inset to Fig.~\ref{Fig.2}(b) and Fig.~S2 in {\it \textbf{Supporting Information (SI)}}]. The fitted $A$ coefficient for $p$=3.80 GPa is 0.302 $\mu\Omega$$\cdot$cm/K$^2$. From the Kadowaki-Woods ratio for a Kramers doublet ground state\cite{KW,Tsujii-KW}, this $A$ coefficient implies a Sommerfeld coefficient $\gamma$=170 mJ/(mol$\cdot$K$^2$), a value nearly three times that at ambient pressure [$\sim$65mJ/(mol$\cdot$K$^2$)]\cite{Luo-CeNi2As2}. We also point out that the smaller cell volume, isostructural analog CeNi$_2$P$_2$ is an intermediate valence compound\cite{Suzuki-CeNi2X2}. Our hydrostatic pressure experiment on CeNi$_{2-\delta}$As$_2$ is qualitatively consistent with a chemical pressure effect induced by P/As doping\cite{Luo-CeNiAsPO}.

\begin{figure}
\includegraphics[width=8.5cm]{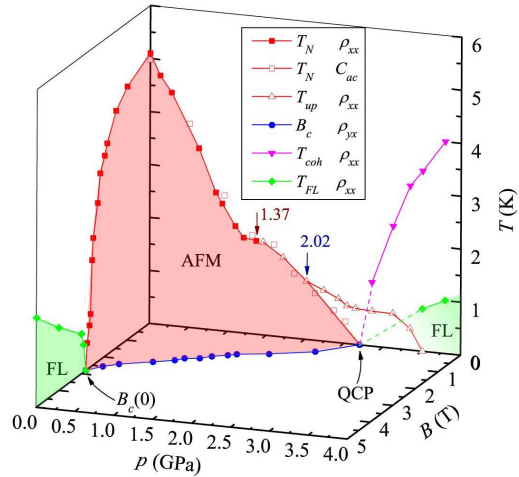}
\caption{The $B$-$p$-$T$ phase diagram of CeNi$_{2-\delta}$As$_2$.
For clarity, we plot $B_c(p)$ in the $T$=0 plane instead of the $T$=0.3 K plane where measurements were made.}\label{Fig.3}
\end{figure}

Figures~\ref{Fig.2}(a-b) summarize the effect of pressure on the field-dependent Hall resistivity $\rho_{yx}(B)$ at $T$=0.3 K. The exchange interaction among Ce moments serves as an ``effective internal field'' that produces an AHE in addition to the normal Hall effect induced by a
Lorentz force. For example, in Fig.~\ref{Fig.2}(c) we show $\rho_{yx}(B)$ at $T$=2 K under ambient pressure. The step-like increase in $\rho_{yx}(B)$ near $B$=2.55 T is reminiscent of the spin-flop transition\cite{Luo-CeNi2As2} observed in isothermal magnetization $M_c(B)$ plotted in the inset to Fig.~\ref{Fig.2}(c). Indeed, the $\rho_{yx}(B)$ curve can be well fit to the relation\cite{Smith-AHE,Hurd-Hall}
\begin{equation}
\rho_{yx}(B)=R_H B+R_S \mu_0 M_c(B),
\label{Eq.1}
\end{equation}
in which $R_H$ is the normal Hall coefficient and the second term characterizes the AHE contribution. The best fit leads to $R_H$=$-1.58\times$10$^{-8}$ m$^3$/C, and $R_S$=2.8$\times$10$^{-6}$ m$^3$/C. The critical field $B_c$ for the spin-flop transition can therefore be defined from the peak in $d\rho_{yx}/dB$ as depicted in Fig.~\ref{Fig.2}(d).
Obviously, $B_c$ moves to lower fields as $p$ increases and terminates near 2.7 GPa as shown in Fig.~\ref{Fig.3}. Taking a single band approximation for simplicity (which is also the upper bound of the multiband situation), the large magnitude of $R_H$ corresponds to a low carrier density $n_c$=3.94$\times$10$^{20}$ cm$^{-3}$, i.e., there are only $\sim$0.032 conduction electrons per formula unit, which corroborates the semi-metallicity of CeNi$_{2-\delta}$As$_2$. The reason for such a low carrier density in CeNi$_{2-\delta}$As$_2$ is still unclear, but it is possible that Ni deficiency\cite{Luo-CeNi2As2,Suzuki-CeNi2X2} has shifted the Fermi level, leaving only a few carriers in the bottom of the renormalized conduction band. At pressures close to and well above 2.7 GPa, the transverse Hall resistivity remains non-linear in field, reflecting the sum of skew scattering due to strong paramagnetism of Ce moments and a contribution from the normal Hall effect.

\begin{figure}
\includegraphics[width=8.5cm]{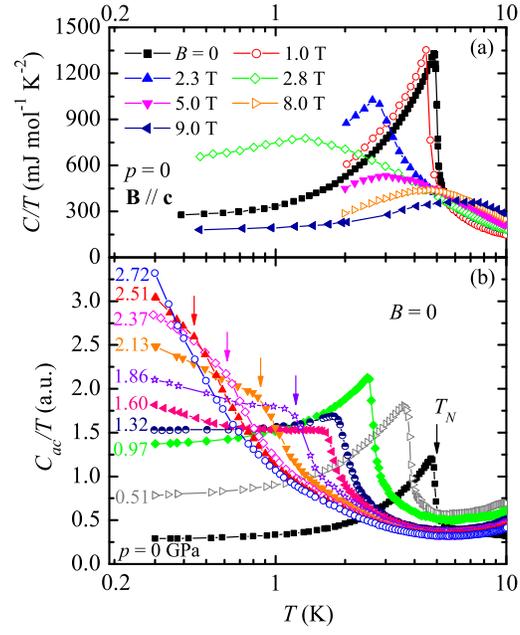}
\caption{(a) Atmospheric pressure specific heat of CeNi$_{2-\delta}$As$_2$ under various magnetic fields, with $\textbf{B}$$\parallel$$\textbf{c}$. (b) Temperature dependent ac heat capacity under pressure; the arrows mark the positions of the AFM transitions.}\label{Fig.4}
\end{figure}

A global $B$-$p$-$T$ phase diagram is plotted in Fig. \ref{Fig.3}. The field dependence of $T_N$ at ambient pressure (on the $B$-$T$ plane) is  derived from  combinations of $\rho_{xx}(T)$ at fixed $B$ and $\rho_{ij}(B)$ at fixed $T$; whereas, $T_N(p)$ is determined from $\rho_{xx}(T)$ and ac heat capacity measurements discussed below. The $B$-$p$ boundary is defined from the pressure dependence of $d\rho_{yx}/dB$. At the zero-pressure critical field, $B_c(0)$=2.8 T, the $T$-linear specific heat is a maximum [$\gamma_0$=657 mJ/(mol$\cdot$K$^2$), as shown in Fig.~\ref{Fig.4}(a)] and $\rho_{xx}(T)$ increases as $T^{1.53}$, indicative of a state similar to that found at field-tuned quantum criticality in metamagnetic systems\cite{Grigera-Sr3ru2O7,Balicas-CeAuAb2,Mun-CeNiGe3,Millis-MMQCP}. {\it \textbf{SI}} provides more details. The $B_c(p)$ line at 0.3 K is continuous and terminates in zero field at $p_c$=2.7 GPa at which point $T_{coh}(p)$ and $T_{FL}(p)$, defined by data such as plotted in the inset to Fig.~\ref{Fig.2}(b), approach the $T$=0 (0.3 K) plane. Combined with the recovery of FL-like behavior and the enhancement of quasiparticle effective mass, $p_c$ defines a zero-field magnetic QCP. The question, though, is what is the nature of the two quantum criticalities, one in zero pressure at 2.8 T and the other in zero field at 2.7 GPa.

The presence of a spin-flop transition and the high magnetic anisotropy\cite{Luo-CeNi2As2} at atmospheric pressure suggests that the magnetic order is of a local-moment type. On the other hand, this spin-flop transition, driven by a
magnetic field, lacks spontaneous time reversal symmetry breaking at the critical field: before the system reaches an intrinsic paramagnetic state, the moments already have been polarized [Fig.~\ref{Fig.4}(a)]. And moreover, considering the weakly first-order nature of this field-induced spin-flop transition even at the low temperature of 0.3 K [see Fig.~S1(c) in {\it \textbf{SI}}], $B_c$=2.8 T is probably very close to a QCP.

In the absence of magnetic field, the pressure-induced quantum-phase transition at $p_c$=2.7 GPa should be a second-order QCP. To further address this, we show ac heat capacity data ($C_{ac}/T$) in Fig.~\ref{Fig.4}(b). It is clearly seen that the $\lambda$-shaped peak corresponding to the AFM transition is gradually suppressed by pressure, and becomes undetectable at 2.72 GPa. We should also note that $C_{ac}/T$ at this pressure roughly obeys a $-\log T$ law at low temperature, strongly demonstrating a NFL behavior with divergent Sommerfeld coefficient that is in contrast to the one induced by field at $B_c$=2.8 T [Fig.~\ref{Fig.4}(a), and {\it \textbf{SI}}].

\section{Discussion}

The disparity between $B$- and $p$-induced quantum criticalities is reminiscent of CeCu$_{6-x}$Au$_x$ in which Au-doping induces local-moment-like antiferromagnetism for $x$$>$0.1 and non-SDW criticality, yet the field-induced critical behavior is characteristic of a three-dimensional SDW QCP\cite{HvL2001}. In this regard, it is interesting that the nominally isoelectronic Au substitution for Cu results in a large reduction in carrier concentration, with $n_c$=$-$0.73/f.u. for $x$=0 and $+$0.061/f.u. for $x$=0.2, which is accompanied by a nearly five-fold increase in the low-temperature resistivity\cite{Bartolf,HvL1998}. This change is not due to the emergence of AFM\cite{Bartolf}. An apparent dichotomy in the nature of the $T$=0 boundaries as a function of doping (or pressure) versus that of field in CeCu$_{6-x}$Au$_x$ is found as well in CeNi$_{2-\delta}$As$_2$.

On the $B_c(p)$ line connecting $B_c$ and $p_c$, there is a slight ``bump'' beginning at 1.37 GPa where a ``bump'' also appears on the line of $T_N(p)$. This pressure also coincides with a change in the resistive signature for magnetic order [Figs.~\ref{Fig.1}(b) and (c)] where at 1.37 GPa $\rho_{xx}(T)$ turns up through an ``inflection'' at $T_{up}$ as it does in CeCu$_{5.8}$Au$_{0.2}$\cite{HvL2001}. In the case of CeCu$_{5.8}$Au$_{0.2}$, an upturn in resistivity at the AFM transition was attributed to current flow with a component along the ordering wavevector $\textbf{Q}$; whereas, the resistivity turns down below $T_N$ when current flow is perpendicular to $\textbf{Q}$\cite{HvL1998,HvL-CeCuAuR}. This provides a possible explanation for the pressure-induced evolution of the resistive signature for AFM in CeNi$_{2-\delta}$As$_{2}$, namely that modest pressure induces a change in ordering wavevector for $p$$>$1.37 GPa. By measuring magnetoresistivity in the configurations of $\textbf{B}$$\parallel$$\textbf{c}$ and $\textbf{B}$$\parallel$$\textbf{ab}$ (Fig.~S5 in {\it \textbf{SI}}), we indeed find evidence for a magnetic order change from the $\textbf{c}$-axis being the easy axis at low pressures to the $\textbf{ab}$-plane being an easy plane at moderate pressures. Whether $\textbf{Q}$ also changes simultaneously with pressure needs to be clarified in the future by microscopic techniques. This signature for AFM order persists to 2.02 GPa above which $\rho_{xx}(T)$ evolves smoothly from above to below $T_{up}$ [Fig.~\ref{Fig.1}(c)]. Fig.~S2(a) provides a closer look of resistivity upturn in this pressure region. An extrapolation of $T_{up}(p)$ for $p$$\leq$2.02 GPa to $T$=0 intersects the pressure axis at $p_c$ (Fig.~\ref{Fig.3}), providing additional evidence for a pressure-tuned QCP at $p_c$. Nevertheless, above 2.02 GPa, $T_{up}(p)$ deviates from this extrapolation and forms an extra dome ranging from 2.02 GPa to 3.51 GPa that we interpret to be a consequence of the low carrier concentration (see below). In this pressure range $T_{up}(p)$ is not a thermodynamic phase boundary.

Pressure, in general, promotes $f$-$c$ hybridization in Ce-Kondo lattices and eventually suppresses AFM order as Kondo compensation of Ce moments begins to dominate. In CeNi$_{2-\delta}$As$_2$, pressure also appears to reduce even further the low carrier density, as witnessed by an overall increase in $\rho_{xx}(T, p)$ and the variation in $\rho_{yx}(B, p)$, and this counters the tendency for stronger hybridization by decreasing the Kondo-impurity temperature scale and, even more dramatically, the Kondo-lattice temperature scale below which a heavy Fermion liquid develops\cite{Pruschke}. A result of this protracted Kondo screening, which is related to Nozi\`{e}res exhaustion idea that insufficient conduction states are available to screen all the moments in a Kondo lattice\cite{Nozieres1998}, is the stabilization of RKKY-driven AFM order\cite{Pruschke}, which could include a change in ordering wavevector and provide a reasonable interpretation of the origin of a ``bump'' in $T_N(p)$ and $B_c(p)$ lines near 1.37 GPa. This concept of protracted screening also provides an appropriate explanation for the deviation of $T_{up}(p)$ from an extrapolation above 2.02 GPa. In this pressure regime, $T_{up}(p)$ reflects the temperature below which already strong scattering in the incoherent Kondo lattice is enhanced by the proliferation of magnetic fluctuations emerging from the projected QCP at $p_c$ established by specific heat measurements. The evolution of conventional resistive signatures of quantum criticality is masked by this scattering. At pressures sufficiently above $p_c$, signatures of a coherent Kondo lattice ($T_{coh}$ and $T_{FL}$) begin to appear, signalling that Kondo hybridization finally has overcome the counter tendency due to a reduced carrier density.

We cannot discount the possible role of Ni deficiencies and the associated disorder-induced Griffiths-phase singularities, but clearly Kondo physics in a low carrier system is primarily responsible for the $B$-$p$-$T$ phase diagram of CeNi$_{2-\delta}$As$_2$. We note that the specific heat anomaly at $T_N$ is quite sharp [Fig.~4(a)], demonstrating a well-defined second-order phase transition, and that the deficiencies reside in the Ni-As conduction layer while the Ce sub-lattice is free of deficiency. Assuming a spherical Fermi surface topology for simplicity and using the carrier density $n_c$=3.94$\times$10$^{20}$ cm$^{-3}$, residual resistivity $\rho_0$=125 $\mu\Omega$ cm, and Sommerfeld coefficient $\gamma$=65 mJ/mol$\cdot$K$^2$ at atmospheric pressure, we estimate a Fermi wave vector $k_F$=0.227 \AA$^{-1}$, effective mass $m^*$=55 $m_0$, and mean free path $l$=189 \AA. Although this mean free path is shorter than in very clean heavy-fermion compounds, it is much longer than the average separation between Ni-site vacancies (2$a$$\approx$8\AA), suggesting that potential scattering by Ni deficiencies does not dominate the magnitude of the resistivity. In CeNi$_{2-\delta}$As$_2$, the RKKY-driven AFM order of Ce moments, partially compensated by protracted Kondo screening, can be tuned by field or pressure to zero-temperature boundaries. Complications of competing pressure-enhanced hybridization and reduced carrier concentration prevent identification of the precise nature of criticality at $p_c$. Nonetheless, the emergence of a signature for Kondo coherence from $p_c$ and a $-\log T$ dependence of $C_{ac}/T$ at $p_c$ suggest a condition unfavorable to SDW criticality but consistent with a local moment type of criticality\cite{HvL,Si-QCP,Luo-CeNiAsPO}. A measure of the Fermi surface evolution around $p_c$ would be instructive.

\section{Conclusion}
To summarize, we have mapped out the global $B$-$p$-$T$ phase diagram of the low carrier density AFM Kondo semi-metal CeNi$_{2-\delta}$As$_2$ by transport and thermodynamic measurements. There are two $T$=0 boundaries on this phase diagram, one induced by magnetic field and the other by pressure. The field-tuned boundary at $B_c$=2.8 T, of weakly first-order nature, is probably very close to a $T$=0 QCP of some type, while the pressure-tuned QCP at $p_c$=2.7 GPa is accompanied by the development of Kondo coherence and divergent quasiparticle effective mass, and thus points to an unconventional QCP. The competition between low carrier density and pressure-enhanced Kondo hybridization plays an important role in the evolution of N\'{e}el order and signatures of criticality. CeNi$_{2-\delta}$As$_2$ provides an interesting paradigm of quantum criticality in the limit of low carrier density, evoking the need for a reexamination of Nozi\`{e}res exhaustion problem and its possible consequences for quantum criticality.



\begin{acknowledgments}
We thank E. D. Bauer, R. Movshovich, and M. Janoschek for helpful discussions.Work at Los Alamos was performed under the auspices of the US Department of Energy, Division of Materials Sciences and Engineering. Y.L. acknowledges a Director¡¯s Postdoctoral Fellowship supported through the Los Alamos Laboratory Directed Research and Development (LDRD) program. Work at Zhejiang University was supported by the National Science Foundation of China (Grants 11374257 and 11190023) and the Fundamental Research Funds for the Central Universities. T.P. acknowledges support from a National Research Foundation (NRF) grant funded by the Ministry of Science, Information and Communications Technology (ICT) and Future Planning of Korea (no. 2012R1A3A2048816).\\
\end{acknowledgments}
\emph{}

\textbf{Materials and Methods}\\
Millimeter sized single crystals of ThCr$_2$Si$_2$-type CeNi$_{2-\delta}$As$_2$ were synthesized by a NaAs-flux method as described elsewhere\cite{Luo-CeNi2As2}. Rietveld analysis of X-ray spectra obtained on powdered single crystals confirms the I4/$mmm$ structure and indicates that the Ni site occupancy is 0.856, close to the result of 0.86 obtained from energy-dispersive x-ray (EDX) microanalysis measurements. 
Electrical transport measurements were made on two samples (noted by \textbf{S1} and \textbf{S2}) as functions of pressure and field. \textbf{S1} was pressurized in an indenter-type cell up to 3.80 GPa in the configuration \textbf{B}$\parallel$\textbf{c}; whereas, measurements on \textbf{S2} were performed in a piston-clamp cell up to 2.65 GPa with \textbf{B} perpendicular to \textbf{c}. Data on both samples agree quantitatively. Ohmic contacts were made in a Hall-bar geometry, and in-plane electrical resistivity ($\rho_{xx}$) and Hall resistivity ($\rho_{yx}$, \textbf{S1} only) down to 0.3 K were measured by an LR-700 ac resistance bridge. Heat capacity of CeNi$_{2-\delta}$As$_2$ under pressure (up to 2.72 GPa) was measured by an ac calorimetric method. For all these measurements, Daphne oil 7474 was employed as a pressure-transmitting medium, and the pressure was determined by measuring the superconducting transition of Pb.\\

\textbf{Author contributions}\\
Y.L. and J.D.T. designed research; Y.L., F.R., and N.W. performed research; Y.L., X.L., T.P., and Z.-A.X. contributed new reagents/analytic tools; Y.L., F.R., and J.D.T. analyzed data; and Y.L., F.R., and J.D.T. wrote the paper.\\

\textbf{Author information}\\
The authors declare no competing financial interests. Correspondence and requests for materials should be addressed to Yongkang Luo (ykluo@lanl.gov) or Joe D. Thompson (jdt@lanl.gov).\\

\newpage
\textbf{Figure legend:}

\newpage
\setcounter{table}{0}
\setcounter{figure}{0}
\setcounter{section}{0}
\onecolumngrid

\emph{}\\
\emph{}\\
\emph{}\\
\emph{}\\
\emph{}\\
\emph{}\\
\emph{}\\
\emph{}\\
\emph{}\\
\emph{}\\
\emph{}\\
\emph{}\\
\newpage

\begin{center}
{\bf \Large
{\it Supporting Information:}\\
Pressure-tuned quantum criticality in the antiferromagnetic Kondo semi-metal CeNi$_{2-\delta}$As$_2$
}
\end{center}

\begin{center}
Yongkang Luo$^{1*}$, F. Ronning$^{1}$, N. Wakeham$^{1}$, Xin Lu$^{2}$, Tuson Park$^{3}$, Zhu-an Xu$^{4}$, and J. D. Thompson$^{1}$ \\
$^1${\it Los Alamos National Laboratory, Los Alamos, New Mexico 87545, USA;}\\
$^2${\it Center for Correlated Matter, Zhejiang University, Hangzhou 310058, China;}\\
$^3${\it Department of Physics, Sungkyunkwan University, Suwon 440-746, South Korea;}\\
$^4${\it Department of Physics, Zhejiang University, Hangzhou 310027, China.}

\end{center}

\renewcommand{\thefigure}{S\arabic{figure}}
\renewcommand{\thetable}{S\arabic{table}}
In this {\it \textbf{Supporting Information (SI)}}, we provide additional magnetization, resistivity, magnetoresistivity, Hall effect, and specific heat data that further support the discussion and conclusions of the main text.

\section{\textbf{Magnetic susceptibility}}


\vspace*{-15pt}
\begin{figure*}[htbp]
\begin{center}
\hspace*{-25pt}
\includegraphics[width=18.0cm]{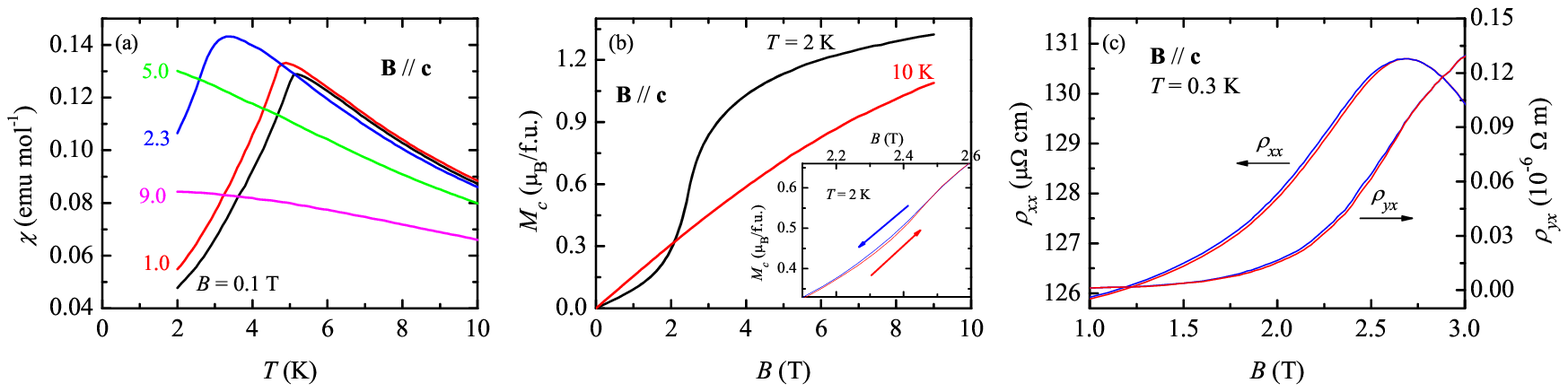}
\vspace*{-25pt}
\caption{\label{FigS1} {\it Magnetic properties of CeNi$_{2-\delta}$As$_2$ at atmospheric pressure. (a) Temperature dependence of the magnetic susceptibility measured at different fields. (b) Isothermal magnetization at $T$=2 K and 10 K. The inset to (b) shows a tiny hysteresis loop near the spin-flop transition. (c) The small hysteresis loops observed in $\rho_{xx}(B)$ and $\rho_{yx}(B)$ at 0.3 K.}}
\end{center}
\end{figure*}

Fig.~S1(a) displays the magnetic susceptibility as a function of temperature with the magnetic field orientation $\textbf{B}$$\parallel$$\textbf{c}$, which is the magnetic easy axis. For a small field $B$=0.1 T, $\chi(T)$ shows a pronounced peak at $T_N$=5.1 K. With increasing magnetic field, this peak in $\chi(T)$ shifts towards lower temperatures and is gradually suppressed. For even higher fields, $\chi(T)$ saturates at low temperature, reflecting spin-polarization of Ce moments. In Fig.~S1(b), we show the isothermal field-dependent magnetization $M(H)$ for temperatures both below and above $T_N$. For $T$$<$$T_N$, $M(H)$ undergoes a metamagnetic (MM) transition near 2.4 T, i.e., $M(H)$ increases linearly below 2.0 T, is followed by a rapid step-like increase near 2.4 T, and finally tends to saturate at high field. This transition is weakly first-order as evidenced by the tiny hysteresis loop in the inset to Fig.~S1(b). We also point out that this weak first-order nature of the MM transition is evident in transport measurements at 0.3 K, as seen in Fig.~S1(c).

\section{\textbf{Electrical transport}}

\vspace*{-15pt}
\begin{figure*}[htbp]
\begin{center}
\hspace*{-25pt}
\includegraphics[width=18.0cm]{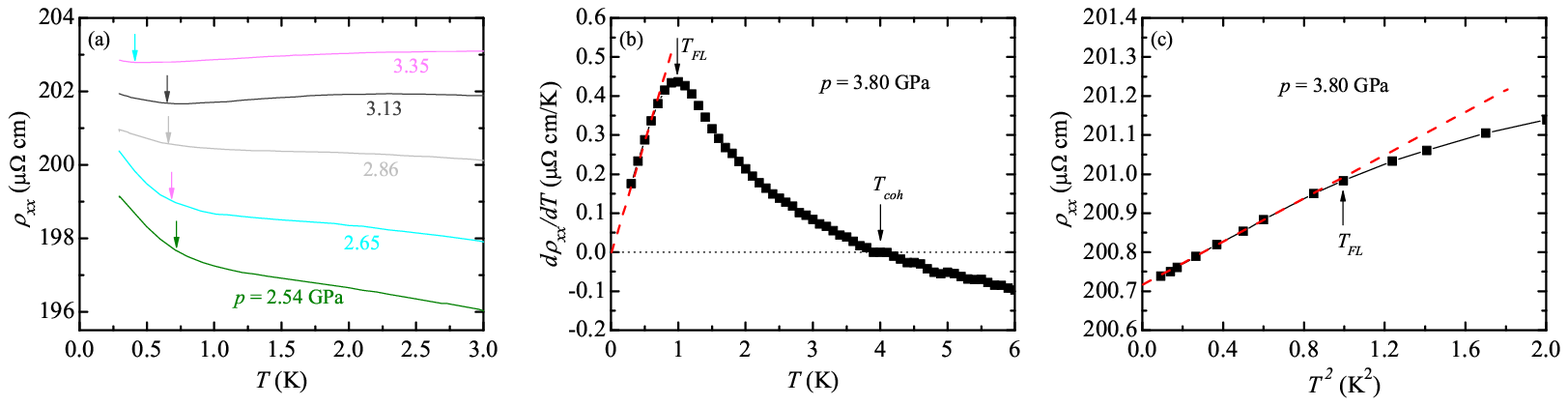}
\vspace*{-25pt}
\caption{\label{FigS2} {\it Temperature-dependent resistivity at high pressure. (a) Enlarged $\rho_{xx}(T)$ at low temperature for $p$$\geq$2.54 GPa. The curves have been plotted with a constant offset for clarity. The arrows signify $T_{up}$. (b) $d\rho_{xx}/dT$ as a function of $T$ for 3.80 GPa. The data have been interpolated and smoothed by 3-point smoothing. (c) $\rho_{xx}$ versus $T^2$ at 3.80 GPa.}}
\end{center}
\end{figure*}

To have a closer look of the pressure dependent resistive upturn, in Fig.~S2(a) we show $\rho_{xx}(T)$ in the low temperature region for the pressure window ranging from 2.54 GPa to 3.35 GPa. For clarity, the curves have been shifted vertically. The arrows signify the position of $T_{up}$. This upturn in $\rho_{xx}(T)$ is gradually suppressed by pressure, and becomes weak at 3.35 GPa. Also seen from Fig.~S2(a) is the pressure dependent resistivity for $T$$>$$T_{up}$ changes slope from negative to positive, and hence a minimum observed at $T_{up}$, characteristic of the formation of a coherent Kondo lattice.

In Fig.~S2(b-c), we present complementary evidence for a possible Fermi liquid (FL) behavior at $p$=3.80 GPa. Fig.~S2(b) plots $d\rho_{xx}/dT$ as a function of $T$. Presuming $\rho_{xx}=\rho_0+AT^n$ ($\rho_0$ is residual resistivity), the advantage of $d\rho_{xx}/dT$ is that the slope is independent of the value of $\rho_0$. Indeed, $d\rho/dT$ linearly approaches the origin, strongly demonstrating the $T^2$-dependence of $\rho_{xx}(T)$ at low temperature. In Fig.~S2(c), we also show $\rho_{xx}(T)$ versus $T^2$. The linear dependence below 1.0 K provides additional evidence for $\rho_{xx}$$\sim$$T^2$. These data confirm a $T^2$ dependence of $\rho_{xx}$ at low temperature, albeit with large residual resistivity and the short temperature fitting range, and therefore, FL behavior is likely to be recovered. We noticed that similar phenomena were also observed in CeCu$_{5.8}$Au$_{0.2}$, another Kondo compound with low carrier density [19].

\vspace*{-15pt}
\begin{figure*}[htbp]
\begin{center}
\hspace*{-25pt}
\includegraphics[width=18.0cm]{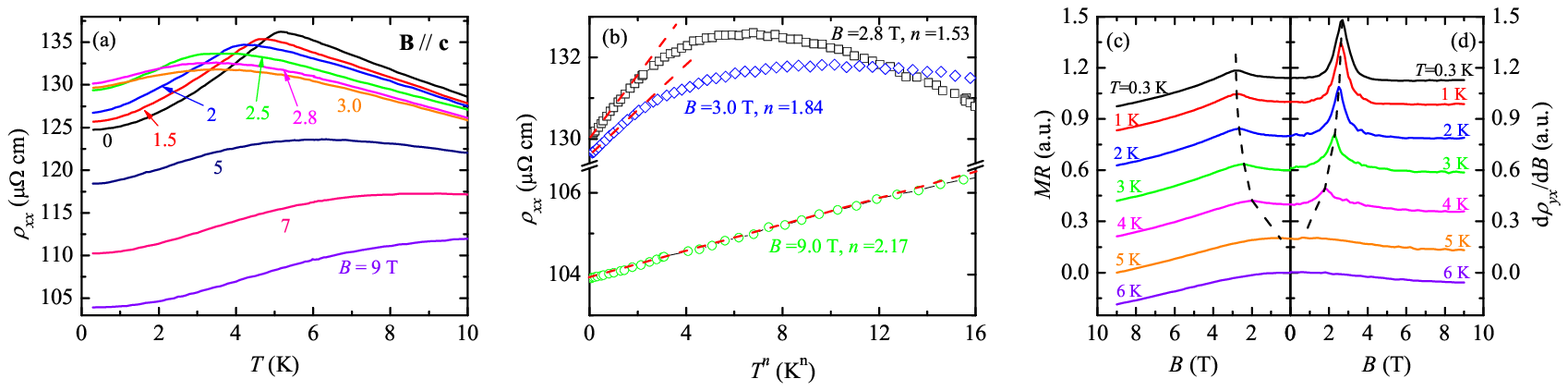}
\vspace*{-25pt}
\caption{\label{FigS3} {\it Field-dependent transport measurements on CeNi$_{2-\delta}$As$_2$ at ambient pressure. (a) Temperature dependence of resistivity under various fields. (b) Low temperature resistivity at $B$=2.8 T, 3.0 T and 9.0 T plotted against $T^n$. Isothermal magnetoresistivity (c) and derivative of the Hall resistivity (d) as functions of $B$. The dashed lines track the critical field $B_c$ as a function of $T$. For clarity, the curves have been shifted vertically.}}
\end{center}
\end{figure*}

In Fig.~S3(a), we present temperature dependencies of resistivity under various fields at atmospheric pressure. For $B$=0, $\rho_{xx}(T)$ exhibits a pronounced peak at $T_N$=5.1 K, a signature of the reduction of spin-scattering off the Ce moments. Increasing external magnetic field gradually depresses this peak, consistent with the magnetic susceptibility [Fig.~S1(a)]. For $B$=2.8 T, which is defined as the critical field of the antiferromagnetic to polarized paramagnetic states (see below), $\rho_{xx}(T)$  has a $T^{1.53}$ dependence, and at even higher fields a FL resistivity, i.e., $\rho_{xx}$=$\rho_0$+$AT^n$ with $n$$\approx$2 (taken as 1.8$\leq$$n$$\leq$2.2) can be observed in a narrow $T$ range. The suppression of AFM order under field can also be found in the isothermal magnetoresistivity [$MR$=$(\rho_{xx}(B)$$-$$\rho_{xx}(0))/\rho_{xx}(0)$] and Hall resistivity ($d\rho_{yx}/dB$) measurements, as seen in  Fig.~S3(c-d). Note that in this context the $MR$ is mainly determined by the spin-scattering by Ce moments, while $d\rho_{yx}/dB$ $\propto$ $ dM_c/dB$ (see main text). The peaks in $MR$ and $d\rho_{yx}/dB$ thus track the critical field as a function of $T$. We summarize the $B$-$T$ phase diagram in Fig.~S4(c). The critical field $B_c$=2.8 T is now unambiguously defined as the point where $T_N$ approaches zero.

\section{\textbf{Specific heat}}

The temperature-dependent specific heat divided by temperature is shown in Fig.~S4(a) at various magnetic fields. The pronounced peak near 5 K at $B$=0 signals the AFM transition, which is gradually suppressed by application of magnetic field. For $B$$\geq$2.8 T, $C/T$ exhibits a Schottky-like broad peak that moves towards higher temperatures as $B$ increases, implying the entrance to a spin-polarized paramagnetic phase. We point out that $C/T$ at low temperatures and at the critical field cannot be fit to either $-\log T$- or $\gamma_0-a T^{0.5}$-dependencies [25, 26, 27]. Instead, $C/T$ at this field tends to crossover to a constant value when $T$$\rightarrow$0. We also measured field-dependent isothermal $C/T$ at $T$=0.45 K, and the results are shown in Fig.~S4(b). A peak centered at $B_c$=2.8 T is clearly seen. The quasiparticle effective mass that can be described by $C/T$ in the $T$$\rightarrow$0 limit, although remaining finite, has increased by more than a factor of 2 at $B_c$ when compared with that at 0 and 9 T. (We note that these low temperature values of $C/T$ are much larger than those estimated by an extrapolation of $C/T$ from above $T_N$ to $T$=0.) Such behavior has been seen in Kondo lattice systems whose MM transition is tuned to $T$=0 by field [16,17], but is substantially different from the case of a second-order quantum phase transition in which the effective mass diverges at the critical field (e.g., the field induced QCP in YbRh$_2$Si$_2$ [28]).

\section{\textbf{$B$-$T$ phase diagram}}

To better understand the state near $B_c$, we systematically analysed the temperature dependence of resistivity by fitting to the formula $\rho_{xx}(T)$=$\rho_0$+$\Delta\rho$=$\rho_0$+$AT^n$. Fig.~S4(d) shows the pressure dependent $\rho_0$ and $n$ in the low temperature limit. Fermi-liquid behavior (1.8$\leq$$n$$\leq$2.2) is restored when $B$$>$$B_c$, while non-Fermi liquid (NFL) behavior ($n$$<$1.8) occurs in the close vicinity of this critical point. Signatures of quantum fluctuations are also reflected in a peak in residual resistivity $\rho_0$ and a dip in the temperature exponent $n$, as shown in Fig.~S4(d). It should be pointed out that $n$ continues to rise as $B$ increases. Similar results were also seen in other MM systems like CeAuSb$_2$ [16] and CeNiGe$_3$ [17]. In CeAuSb$_2$ [16], an unusual scattering mechanism with $T^3$ dependence of $\Delta\rho(T)$ was identified when the magnetic field is high enough and was ascribed to spin scattering off the polarized Ce moments. Therefore, the observed FL region in CeNi$_{2-\delta}$As$_2$ is likely a crossover between NFL and high-$n$ scattering regimes.

\vspace*{-20pt}
\begin{figure*}[htbp]
\begin{center}
\includegraphics[width=17.0cm]{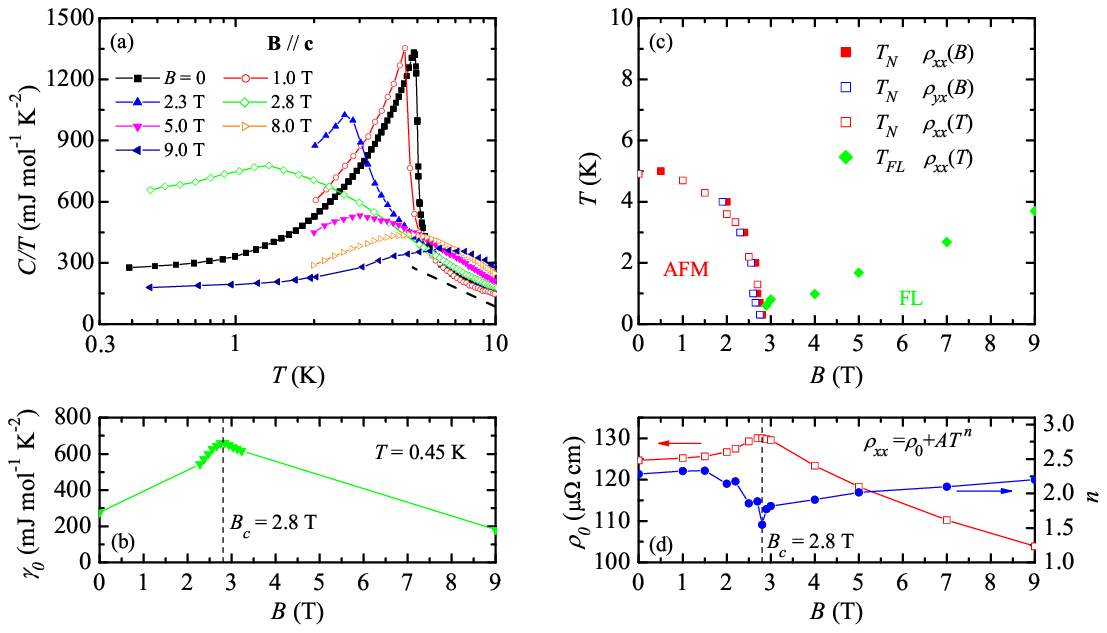}
\vspace*{-40pt}
\caption{\label{FigS4} {\it Evidence for a critcal magnetic field at $p$=0. (a) Temperature dependence of specific heat divided by temperature measured at various fields. The dashed line is a guide to the eyes of the $-\log T$ dependence in the paramagnetic state.  (b) Sommerfeld coeffcient $\gamma_0$ as a function of $B$ measured at $T$=0.45 K. (c) Phase diagram of CeNi$_{2-\delta}$As$_2$ on the $B$-$T$ plane. (d) Plot of the field-dependent residual resistivity $\rho_0$ and the exponent $n$ from fits to the resistivity in the low temperature limit.}}
\end{center}
\end{figure*}

Considering all of our observations, and the weakly first-order nature of the field-induced spin-flop transition even at 0.3 K, it is likely that the critical point $B_c$=2.8 T in CeNi$_{2-\delta}$As$_2$ is also very close to a QCP.

\section{\textbf{Change in magnetic structure with pressure}}

\vspace*{-15pt}
\begin{figure*}[htbp]
\begin{center}
\includegraphics[width=14.0cm]{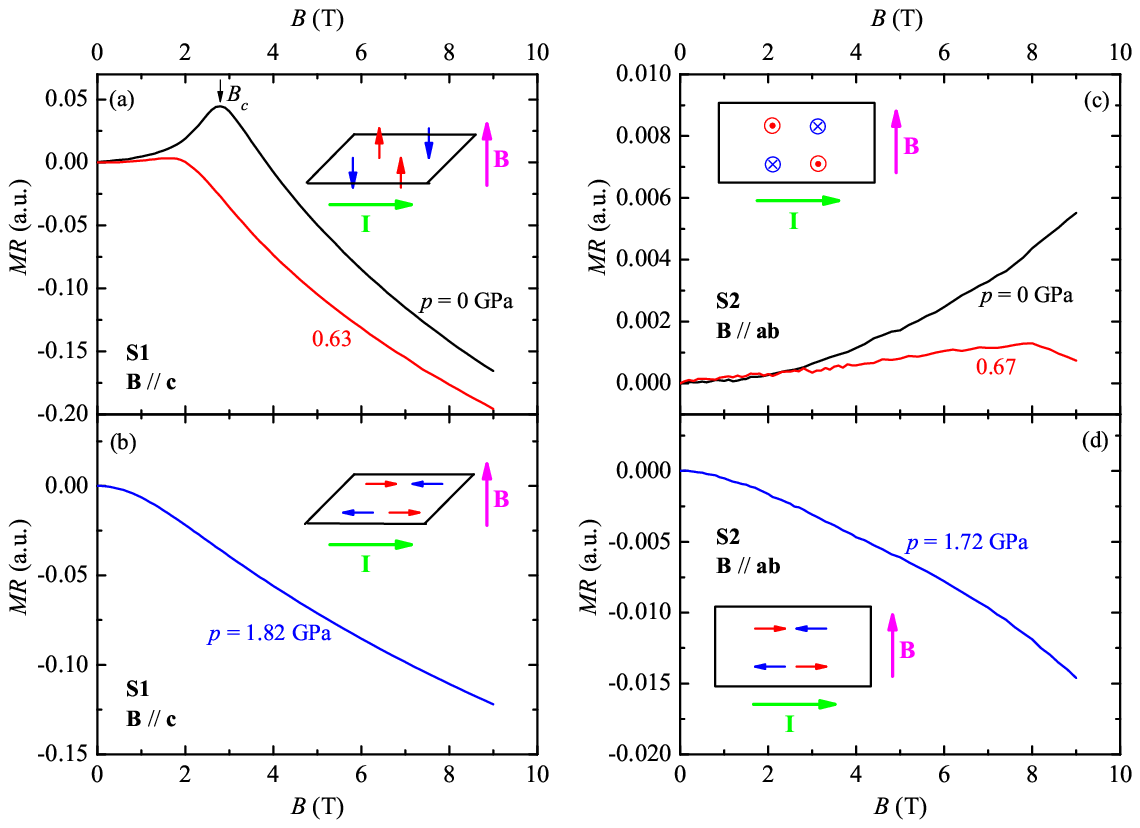}
\caption{\label{FigS5} {\it Magnetoresistivity as a function of $B$. All the data were taken at 0.3 K. (a-b), $MR(B)$ with $\textbf{B}$$\parallel$$\textbf{c}$. (c-d), $MR(B)$ with $\textbf{B}$$\parallel$$\textbf{ab}$. The insets show the schematic diagrams of the geometry between Ce moments and magnetic field: (a) and (c) for ambient pressure, whereas (b) and (d) for pressures $p$$>$1.37 GPa.}}
\end{center}
\end{figure*}

Magnetoresistivity measurements provide evidence for a gradual change in the magnetic structure as a function of pressure.  Figure S5 shows representative plots of magnetoresistivity versus field at various pressures, and in each of the panels, \textbf{B} is perpendicular to the current direction, which is in the \textbf{ab}-plane. In panels (a) and (b), \textbf{B} is along (001), the easy magnetic axis; whereas, in panels (c) and (d), \textbf{B} is perpendicular to (001). In the low field limit, where CeNi$_{2-\delta}$As$_2$ is in its AFM state, the $MR$ is positive for both field directions [panels (a) and (c)] at atmospheric pressure and small applied pressure. In contrast, when pressure is increased to around 1.7-1.8 GPa [panels (b) and (d)], $MR$ in the antiferromagnetically ordered state at low fields is negative irrespective of the direction in which \textbf{B} is applied. These results suggest that the magnetic order has changed from the \textbf{c}-axis being the easy axis at low pressures to the \textbf{ab}-plane being an easy plane at moderate pressures. Whether the magnetic ordering wavevector $\textbf{Q}$ also changes simultaneously is still an open question and needs to be clarified by microscopic techniques.

\end{document}